\documentclass[%
 reprint,
 amsmath,amssymb,
 aps,
 pre,
]{revtex4-1}

\usepackage{mathrsfs} 

\usepackage{graphicx}
\usepackage{dcolumn}
\usepackage{bm}
\usepackage[hidelinks]{hyperref}

\usepackage[table,xcdraw]{xcolor}
\usepackage{adjustbox}

\usepackage[all]{background}
\usepackage{url}

\SetBgContents{\href{https://doi.org/10.1103/PhysRevE.95.042608}{Published at Phys. Rev. E \textbf{95}, 042608 (24 April 2017)}}
\SetBgPosition{3.5cm,0.7cm}
\SetBgOpacity{0.5}
\SetBgAngle{0.0}
\SetBgScale{2.0}

\begin{document}


\title{Active rotational and translational microrheology beyond the linear spring regime}

\author{Lachlan J. Gibson}
\author{Shu Zhang}%
\author{Alexander B. Stilgoe}
\author{Timo A. Nieminen}
\author{Halina Rubinsztein-Dunlop}
\affiliation{%
The University of Queensland, School of Mathematics and Physics, Brisbane QLD 4072, Australia
}%

\date{\today}

\begin{abstract}
Active particle tracking microrheometers have the potential to perform accurate broad-band measurements of viscoelasticity within microscopic systems. Generally, their largest possible precision is limited by Brownian motion and low frequency changes to the system. The signal to noise ratio is usually improved by increasing the size of the driven motion compared to the Brownian as well as averaging over repeated measurements. New theory is presented here whereby error in measurements of the complex shear modulus can be significantly reduced by analysing the motion of a spherical particle driven by non-linear forces. In some scenarios error can be further reduced by applying a variable transformation which linearises the equation of motion. This enables normalisation that eliminates error introduced by low frequency drift in the particle's equilibrium position. Our measurements indicate that this can further resolve an additional decade of viscoelasticity at high frequencies. Using this method will easily increase the signal strength enough to significantly reduce the measurement time for the same error. Thus the method is more conducive to measuring viscoelasticity in slowly changing microscopic systems, such as a living cell.
\end{abstract}

\pacs{Valid PACS appear here}
\maketitle


\section{Introduction}
The strength of a microrheometer can be assessed by its ability to perform accurate broad-band measurements of viscoelasticity within microscopic systems. In particular, there is great interest in improving methods for conducting measurements within living biological systems, such as a cell\cite{Weihs,Robert,Kollmannsberger,Rogers,Chen}.

Particle tracking microrheometers have proven to be a good candidate for accomplishing such a task\citep{Rogers,Chen,Wirtz}. They work by tracking the motion of one or more particles embedded in the pertinent medium. The complex shear modulus $G^*(\omega)$, a frequency ($\omega$) dependent measure of linear viscoelasticity, can be inferred from the way the particles move\cite{MasonWeitz,Bennett,Preece}.

Biological systems are often very small or highly inhomogeneous\cite{Tseng2}. So, tracking only a single particle can be more practical since the measurement is more localised than a multi-particle system. The motion of a single tracked particle can either be driven passively, where Brownian motion is the primary driving force, or actively, where Brownian motion acts as a noise on top of another external driving force. For example, Bennett \textit{et al}.\cite{Bennett} trapped a single spherical birefringent particle using optical tweezers. The particle's birefringence also allowed it to be angularly trapped when using a linearly polarised laser beam. In this particular example, the angular motion driven by thermal fluctuations allowed $G^*(\omega)$ to be calculated using statistical methods including autocorrelations. Therefore, passive methods tend to be more successful at measuring higher frequency viscoelasticity. Conversely, passive methods require too much time to resolve lower frequency viscoelasticty precisely\citep{Preece,Tassieri} in slowly changing systems\cite{Tseng1,Tassieri2015}.

Active methods, in which the particle is driven by some other force, often eliminate Brownian noise by averaging over a repeated motion. The average Brownian motion reduces towards zero leaving only the nonstochastic motion. For example, Preece, \textit{et al}.\cite{Preece} used optical tweezers to trap a spherical particle within two alternating spatially offset traps. The particle switched between one stable equilibrium to another when one beam was turned off and the other turned on. The linear motion of the particle as it fell into each trap was measured and used to calculate $G^*(\omega)$.

Evidently, it is possible to measure viscoelasticity by examining either rotational or linear motion. Therefore, the aim of this paper is to outline and test a generalised theory applicable to either kind of motion. This theory describes how to obtain $G^*(\omega)$ from repeated measurements of a particle falling into an equilibrium position under the influence of both Brownian noise and a position dependent force.

For the sake of simplicity, the previous theory (such as that used by Preece, \textit{et al}.) assumed a force that is linearly dependent on position. For small displacements this is often a valid assumption. However, as will be subsequently shown in section \ref{sec_Error_Analysis}, the signal strength can be significantly increased by allowing the particle to fall into position from outside the linear regime. Increasing the signal strength of each individual measurement can appreciably reduce the total measurement time, thereby justifying application of this method in dynamic biological systems such as a living cell. Therefore, the theory outlined here  accounts for non-linear driving forces (not to be confused with non-linear motion or non-linear viscoelasticity).

To confirm the validity of the theory in at least one example, experimental measurements in both viscous and viscoelastic fluids conducted by an optical tweezers microrheometer are also examined. The different analysis methods for the same data are applied to compare the accuracy as well as the frequency range in which the viscoelasticity can be resolved.

It should be stressed that although the theory is only experimentally verified in this paper using optical tweezers measurements, the analysis is not predicated on that mode of particle manipulation. Provided the driving force is characterisable, this theory could also be applied to many other systems such as magnetic or acoustic tweezers.

\section{Theory}
For simplicity, the following theory is expressed in terms of rotational dynamics. However, obtaining the corresponding results for linear motion at any step can be achieved by a simple substitution. Angle, moment of inertia, torque and rotational drag can be replaced by their respective linear counterparts: linear position, mass, force and linear drag.
\subsection{Equation of Motion}
\subsubsection{jth Flip Langevin Equation}
Consider a microscopic spherical particle centred at the origin with a fixed centre of mass. The particle, embedded in a fluid with linear viscoelasticity, is free to rotate about the $z$-axis guided by Brownian motion, viscoelastic drag and an angular dependent driving torque. The particle should have a stable equilibrium angle such that it becomes trapped at a root of the driving torque function. Repeatedly dropping the particle into the trap from an outside position allows the Brownian noise to be mitigated by averaging many drops.

With a moment of inertia $I$ the stochastic evolution of the azimuthal angle ($\phi_j$) of the $j$th drop can be modelled by a generalised Langevin equation\citep{MasonWeitz,Bennett}
\begin{equation}\label{eq_langevin}
I\ddot{\phi_j}=\tau_j(t)-\int_{-\infty}^{t}\zeta(t-t_{l})\dot{\phi}_j(t_{l})\, dt_{l}-\chi T(\phi_j).
\end{equation}

\noindent The total toque ($I\ddot{\phi_j}$) on the sphere at time $t$ is the sum of the driving torque ($-\chi T(\phi_j)$) that forms the trap, the viscoelastic torque ($-\int_{-\infty}^{t}\zeta(t-t_{l})\dot{\phi_j}(t_{l})dt_{l}$ with generalised memory function $\zeta(t)$) from the fluid and the thermal toque ($\tau_j(t)$) from Brownian motion.

\subsubsection{Driving Torque Function Properties}\label{sec_T_properties}
Without loss of generality, the stable equilibrium angle is set to $0$ with positive trap stiffness $\chi$ so that $T(0)=0$ and $T'(0)=1$. In contrast to the dot symbol in equation \ref{eq_langevin} which denoted a time derivative, here the prime symbol indicates a spatial derivative. The trap potential is assumed to be symmetric about the equilibrium whereby the so called driving torque function, $T(\phi)$, is a continuously differentiable odd function. Hence, for small deviations about the equilibrium, the Taylor series of $T(\phi)$ to fifth order is given by

\begin{equation}\label{eq_Taylor}
T(\phi)=\phi+\frac{T_3}{3!} \phi^3+\frac{T_5}{5!} \phi^5+ \dots
\end{equation}

\noindent where $T_n=T^{(n)}(0)$. Notice that, all even terms in the series are zero since $T(\phi)$ is an odd function.

In order for the particle to be pulled into the $\phi=0$ equilibrium the driving torque must have opposite sign to the position. Therefore, the torque function must have the same sign as the position, $\text{sign}(T(\phi))=\text{sign}(\phi)$. This requirement can limit the allowed positions if the torque changes sign. Therefore, if there exists an angle $\phi=R>0$ such that $T(R)=0$ then the domain must be restricted to $|\phi|<R$. Similarly, if there exists a singularity at angle $\phi=R>0$ such that $\lim_{\phi\to R}T(\phi)^{-1}=0$, then the domain is also restricted to $|\phi|<R$. Since this restriction applies to all roots/singularities (except for $\phi=0$) $R$ is chosen to be the smallest positive root/singularity. If $T(\phi)$ has no additional roots to $\phi=0$ and is continuously differentiable over all $\mathbb{R}$, then the domain is unrestricted, $\phi\in\mathbb{R}$.

\subsubsection{Stokes Flow}
Particle tracking microrheometers typically operate with microscopic particles. Therefore, it is likely that the fluid has a low Reynolds number ($\mathscr{R}\ll 1$) and hence undergoes Stokes flow\cite{Bennett}. The inertial term $I\ddot{\phi_j}$ in equation \ref{eq_langevin} is, consequently, negligible relative to the others and can be ignored,

\begin{equation}\label{eq_lowrenolds}
\tau_j(t)=\int_{-\infty}^{t}\zeta(t-t_{l})\dot{\phi}_j(t_{l})\, dt_{l}+\chi T(\phi_j).
\end{equation}

\subsubsection{Generalised Memory Function}
The time dependent generalised memory function, $\zeta(t)$, describes the ratio of viscoelastic torque to an instantaneous step rotation of the particle. Hence, it is proportional to the fluid's relaxation modulus\cite{MasonWeitz},

\begin{equation}
\zeta(t)=\alpha G_r(t),
\end{equation}

\noindent where $\alpha$ depends on the geometry of the probe particle as well as the type of motion. Therefore, the Langevin equation relates the fluid viscoelastiticy to the angular position by

\begin{equation}\label{eq_langevinstokeseinstein}
\tau(t)=\alpha\int_{-\infty}^{t}G_r(t-t_{l})\dot{\phi}_j(t_{l})\, dt_{l}+\chi T(\phi_j).
\end{equation}

\noindent For a sphere of radius $a$ undergoing rotational or linear motion
\begin{equation}
\alpha=8\pi a^3,\, \text{or } \alpha=6\pi a
\end{equation}
respectively\cite{lamb1895hydrodynamics}.

\subsection{Linear Case}\label{sec_theory_linear_case}
\subsubsection{Normalisation}
If $T(\phi)$ is a non-linear function, then the Langevin equation (\ref{eq_langevinstokeseinstein}) is a non-linear differential equation. This poses a problem for any repeated measurements in which the initial position of each flip varies. If the Langevin equation were linear then the position could be normalised by dividing the equation by the initial angle.

Previously, to obtain a linear differential equation the flipping angle was assumed to be small such that the driving torque could be approximated by its Taylor series (equation \ref{eq_Taylor}) to 1st order,
\begin{equation}
T(\phi)\approx\phi.
\end{equation}

\noindent In this case, transforming to the normalised angle $\varphi_j=\frac{\phi_j}{\phi_j(0)}$ so that $\varphi_j(0)=1$ gives
\begin{equation}\label{eq_norm_linear_Langevin}
\frac{\tau_j(t)}{\phi_j(0)}=\alpha\int_{-\infty}^{t}G_r(t-t_{l})\dot{\varphi}_j(t_{l})\, dt_{l}+\chi \varphi_j.
\end{equation}

Notice that, after normalisation the Brownian motion term is inversely proportional to the initial position $\phi_j(0)$. Therefore, to minimise the effect of Brownian motion the initial angle should be maximised, but only within the allowed domain that satisfies the Taylor series small angle approximation. Thus, there exists some optimal angle whereby the total error contributed by Brownian motion and the Taylor series is minimised. The value of this optimal angle and relative error is quantified later in section \ref{sec_Linear_Error}.

\subsubsection{Average Flip}
The Brownian noise can be reduced by averaging $n$ repeated flips. Assuming each flip is independent of the others, the normalised linear Langevin equations (\ref{eq_norm_linear_Langevin}) for each rotation can be averaged,
\begin{equation}\label{eq_average_linear}
0=\alpha\int_{0}^{t}G_r(t-t_{l})\dot{\varphi}(t_{l})\, dt_{l}+\chi \varphi.
\end{equation}

\noindent $\varphi$ represents the expected normalised angle and is estimated using a finite average of all $n$ flips,

\begin{equation}
\varphi\approx\frac{1}{n}\sum_{j=1}^n \varphi_j.
\end{equation}

Provided that the time between flips is much longer than the time it takes for the particle to reach equilibrium, each flip should `forget' the previous one and finish with an average velocity of zero. Mathematically, this is expressed as $\dot{\varphi}(t)=0$, for $t<0$, which truncates the memory integral at $t=0$. The average Brownian motion is also assumed to be zero, removing the corresponding term entirely.

\subsubsection{Viscous Fluid}
A purely viscous fluid without any elasticity does not `remember' any past motion. Its relaxation modulus is proportional to a Dirac delta function, $G_r(t)=\eta \delta(t)$, where $\eta$ is the dynamic viscosity. With this relaxation modulus, equation \ref{eq_average_linear} simplifies to a simple first order ordinary differential equation,

\begin{equation}\label{eq_viscous_linear}
0=\alpha\eta\dot{\varphi}+\chi \varphi,
\end{equation}

\noindent with a well known solution,

\begin{equation}
\varphi=e^{-k t}, \text{where } k=\frac{\chi}{\alpha\eta}.
\end{equation}

\noindent Evidently, the viscosity is inversely related to the decay rate of the angle over time, $k$.

\subsubsection{Unilateral Fourier Transform}\label{sec_UFT}
More generally, obtaining linear viscoelasticity from the dynamics requires the use of a unilateral Fourier transform (UFT). Represented by a tilde, the UFT is defined by
\begin{equation}
\tilde{f}(\omega)=\int_{0}^{\infty}f(t)e^{-i\omega t}\, dt.
\end{equation}

Applying the UFT to equation \ref{eq_average_linear} transforms the convolution integral into a product that can be easily manipulated,

\begin{equation}\label{eq_transformed_langevin}
0=\alpha \tilde{G}_r(\omega)\left(i\omega\tilde{\varphi}-1\right)+\chi \tilde{\varphi}.
\end{equation}

The relaxation modulus, $G_r(t)$, is related to the time domain conjugate of $G^*(\omega)$ by

\begin{equation}
G^*(\omega)=i\omega\tilde{G}_r(\omega).
\end{equation}

\noindent Therefore, $G^*(\omega)$ can be expressed in terms of $\tilde{\varphi}$ by rearranging equation \ref{eq_transformed_langevin},

\begin{equation}\label{eq_G_linear}
G^*(\omega)=\frac{\chi}{\alpha} \frac{i\omega\tilde{\varphi}}{1-i\omega \tilde{\varphi}}.
\end{equation}

\noindent Equation \ref{eq_G_linear} relates the linear viscoelasticity to the average motion of the particle at different time scales.

\subsection{Non-linear Case}
The theory presented thus far acts mostly as a summary of already known methodology for the purpose of juxtaposition. This section will now adjust the theory to account for a non-linear driving torque function.

\subsubsection{Viscous Case}
Consider the average behaviour given by a non-linear driving torque function in a viscous fluid. The Langevin equation is similar to equation \ref{eq_viscous_linear}, but is a non-linear ordinary differential equation,
\begin{equation}\label{eq_viscous_nonlinear}
0=\alpha\eta\dot{\phi}+\chi T(\phi).
\end{equation}

\noindent Notice the assumption, that

\begin{equation}
T(\phi)\approx\frac{1}{n}\sum_{j=1}^n T(\phi_j),
\end{equation}

\noindent which should be valid provided the deviations from the average of each individual flip are not too large.

\subsubsection{Variable Transform}
The non-linearity of equation \ref{eq_viscous_nonlinear} makes it non-normalisable in terms of $\phi$. However, applying a variable transformation can make it normalisable in terms of a different variable,
\begin{equation}
\Psi(\phi)=\exp\left(\int \frac{d\phi}{T(\phi)}\right).
\end{equation}

\noindent More specifically, the new position variable $\Psi$ is defined as the solution to
\begin{equation}\label{eq_psi_def}
\Psi=T\Psi' , \text{ s.t. } \Psi'(0)=1.
\end{equation}

Applying this transformation linearises equation \ref{eq_viscous_nonlinear},

\begin{equation}
0=\alpha\eta\dot{\Psi}+\chi \Psi,
\end{equation}

\noindent which, like the viscous linear case, has an exponential solution,

\begin{equation}\label{eq_vis_nonlin_sol}
\psi=e^{-k t},\text{ where, }\psi=\frac{\Psi}{\Psi(\phi_0)},\text{ and, }\phi_0=\phi(0).
\end{equation}

\subsubsection{Properties of $\Psi$}
The definition of $\Psi$ in equation \ref{eq_psi_def} ensures that it is a strictly increasing continuously differentiable odd function of $\phi$ over the whole domain.

Its Taylor series is given by
\begin{equation}\label{eq_Psi_Taylor}
\Psi(\phi)=\phi+\frac{-T_3}{2\times 3!}\phi^3+\frac{5T_3^2-T_5}{4\times5!}\phi^5+\dots
\end{equation}

\noindent where the derivatives of $\Psi$ at $\phi=0$ can be expressed in a recursive form as a discrete convolution,
\begin{equation}\label{eq_psi_recursive}
\Psi_n=-n!\sum_{j=1}^{\frac{n-1}{2}}\frac{n-2j}{n-1}\frac{\Psi_{n-2j}}{(n-2j)!}\frac{T_{2j+1}}{(2j+1)!},
\end{equation}

\noindent where $\Psi_n=\Psi^{(n)}(0)$ and $\Psi_1=1$.

Finding the radius of convergence of this Taylor series in general has proven difficult. However, by dividing equation \ref{eq_psi_recursive} by $\Psi_n$ and taking the $n\to\infty$ limit, it can be shown that if $\Psi_i\geq 0$ for all derivatives, then the radius of convergence either covers the whole domain or is at least as large as the radius of convergence of the $T(\phi)$ Taylor series described in section \ref{sec_T_properties}.

\subsubsection{Solution in Terms of $\phi$ by Inverting $\Psi$}
Finding the solution to equation \ref{eq_viscous_nonlinear} in terms of the original position variable, $\phi$, can be achieved by applying the inverse variable transformation to the solution in terms of $\Psi$ given by equation \ref{eq_vis_nonlin_sol},

\begin{equation}\label{eq_vis_nonlin_sol_phi}
\phi=\Psi^{-1}(\Psi)=\Psi^{-1}\left(\Psi(\phi_0)e^{-kt}\right).
\end{equation}

The Taylor series of the inverse function $\Psi^{-1}(\Psi)$ can be found by series reversion\cite{morse1953methods} of equation \ref{eq_Psi_Taylor},

\begin{equation}
\Psi^{-1}(\Psi)=\Psi+\frac{T_3}{2\times 3!}\Psi^3+\frac{5T_3^2+T_5}{4\times5!}\Psi^5+\dots
\end{equation}

Therefore, by applying the Taylor series of both $\Psi$ and $\Psi^{-1}$ to equation \ref{eq_vis_nonlin_sol_phi} the solution to equation \ref{eq_viscous_nonlinear} in terms of time and initial position can be found in a series form,

\begin{equation}
\phi=\phi_0 e^{-kt}-\phi_0^3\left(e^{-kt}-e^{-3kt}\right)\frac{T_3}{2\times 3!}+\dots
\end{equation}

Notice that, the series is always exactly correct at the time bounds $t=0$ and $t\to\infty$ irrespective of the degree it may be truncated. The first term is the solution under the small angle approximation and each successive term adds corrections to the position between the time bounds.

\subsubsection{Unnormalised Analysis Viscoelastic Fluid}\label{sec_UnnormalisedAnalysis}
Now consider the average dynamics of a particle in a viscoelastic fluid driven by a non-linear torque function. Without normalisation equation \ref{eq_langevinstokeseinstein} can be averaged. Similar to equation \ref{eq_average_linear}, the average thermal torque and angular velocity for $t<0$ are zero,

\begin{equation}\label{eq_lagevin_nonlinear}
0=\alpha\int_{0}^{t}G_r(t-t_{l})\dot{\phi}(t_{l})\, dt_{l}+\chi T(\phi).
\end{equation}

Following the steps outlined in section \ref{sec_UFT}, applying the unilateral Fourier transform allows $G^*(\omega)$ to be evaluated,

\begin{equation}
G^*(\omega)=\frac{\chi}{\alpha} \frac{i\omega\tilde{T}}{\phi_0-i\omega \tilde{\phi}}.
\end{equation}

\noindent Notice that, the transform of the torque function is evaluated using its implicit time dependence via $T(\phi(t))$.

This expression has a similar form to equation \ref{eq_G_linear}, however, the non-linearity of $\tilde{T}$ means that it must depend on the initial position $\phi_0$. Therefore, any variation in the initial position due to slow changes in the system or apparatus can introduce error to the calculated result.

\subsubsection{Viscoelastic Fluid With Variable Transformation}\label{sec_NormalisedAnalysis}
Motivated by the successful linearisation in the viscous case, the same variable transform is applied to equation \ref{eq_lowrenolds}, which models the dynamics of the $j$th flip driven by a non-linear driving torque function in a viscoelastic fluid,

\begin{align}
\frac{\tau_j(t)}{f(t)}=\alpha\int_{-\infty}^{t}G_r(t-t_{l})\dot{\psi}_j(t_{l}) \frac{f(t_{l})}{f(t)}\, dt_{l}+\chi \psi_j,\\
\text{where }f(t)=\frac{T(\phi(t))}{\psi(t)}=\frac{\Psi(\phi_0)}{\Psi'(\phi)}.
\end{align}

Now it is assumed that the fluid memory function decays much faster than the time of the flip, so that

\begin{equation}\label{eq_slow_approximation}
\frac{f(t_{l})}{f(t)}\approx 1.
\end{equation}

\noindent Notice that, this condition is exactly met in a viscous fluid which has no `memory'. Conversely, for an elastic solid the memory function never decays to zero, so this assumption would invariably fail. Making the approximation simplifies the Langevin equation to a normalisable form reminiscent of the linear case,

\begin{equation}\label{eq_transform_langevin}
\frac{\tau_j(t)}{f(t)}=\alpha\int_{-\infty}^{t}G_r(t-t_{l})\dot{\psi}_j(t_{l})\, dt_{l}+\chi \psi_j.
\end{equation}

Following the same steps of averaging and transforming outlined in section \ref{sec_theory_linear_case} allows the complex shear modulus to be calculated,

\begin{equation}\label{eq_G_nonlinear}
G^*(\omega)=\frac{\chi}{\alpha} \frac{i\omega\tilde{\psi}}{1-i\omega \tilde{\psi}}.
\end{equation}

Evidently, this expression of $G^*(\omega)$ is very similar to equation \ref{eq_G_linear} where the new normalised position variable $\psi$ has taken over the role of $\varphi$. Notice that, in this case minimising the Brownian motion term involves maximising $f(t)$. Generally this also involves increasing $\phi_0$ however the allowed domain is much larger without the Taylor series small angle approximation. Instead the maximum value is only limited by the slow flip time (relative to the fluid memory function) assumption.

\section{Error Analysis}\label{sec_Error_Analysis}
This section aims to quantify the theoretical relative error of both both the old and new methods of analysis. This can help compare both methods and also determine the optimal initial position which minimises these errors.
\subsection{Linear Case} \label{sec_Linear_Error}
\subsubsection{Error in Complex Shear Modulus}

As outlined in section \ref{sec_theory_linear_case}, maximising the signal to noise ratio involves increasing the initial position. However, since the driving torque function is only approximately linear for small angles, increasing $\phi_0$ too much will introduce systematic errors larger than the random error caused by Brownian motion. To quantify these errors $G^*(\omega)$ is calculated directly from the multiple flip average of equation \ref{eq_langevinstokeseinstein}. Except this time the linear torque and zero mean thermal torque approximations are not imposed, $T(\phi)\ne\phi$ and $\tau\ne0$, where $\tau$ is the average thermal torque,

\begin{equation}
G^*(\omega)=\frac{\chi}{\alpha}\frac{i\omega}{\phi_0-i\omega\tilde{\phi}}\left(\tilde{\phi}+\left(\tilde{T}-\tilde{\phi}\right)-\frac{\tilde{\tau}}{\chi}\right).
\end{equation}

\noindent Therefore, the absolute relative error in $G^*(\omega)$ can be evaluated by

\begin{equation}
\delta G^*_{Lin}=\left|\frac{\tilde{T}-\tilde{\phi}}{\tilde{\phi}}-\frac{\tilde{\tau}}{\chi\tilde{\phi}}\right|.
\end{equation}

\subsubsection{Average Thermal Torque}
The average thermal torque defined by
\begin{equation}
\tau(t)=\frac{1}{n}\sum_{j=1}^n \tau_j(t)
\end{equation}

\noindent only approaches zero as $n\to\infty$. For a finite number of flips the average Brownian motion will still have a thermal torque with standard deviation decaying with $n^{-1/2}$.

Therefore, assuming the thermal torque is white noise, the unilateral Fourier transform of $\tau$ should have a constant magnitude that also decays with $n^{-1/2}$. The phase of $\tilde{\tau}$ at each frequency should be random meaning that the expected real and imaginary parts are both zero. Therefore, in calculating the following expected errors, terms proportional to $\tilde{\tau}$ or the real or imaginary parts of $\tilde{\tau}$ can be ignored. So, the expected relative error in the linear case should be,

\begin{equation}
\delta G^*_{Lin}=\sqrt{\left|\frac{\tilde{T}-\tilde{\phi}}{\tilde{\phi}}\right|^2+\left|\frac{\tilde{\tau}}{\chi\tilde{\phi}}\right|^2}.
\end{equation}

\subsubsection{High Frequency Error}
An expression for the relative error at high frequencies can be found by employing the initial value theorem whereby the unilateral Fourier transform at high frequencies can be asymptotically related to the initial value of the function in the time domain,

\begin{equation}\label{eq_IVT}
\tilde{f}(\omega)\sim\frac{f(0)}{i\omega}.
\end{equation}

Applying the initial value theorem as well as the Taylor series of $T(\phi)$ to third order yields

\begin{equation}
\delta G^*_{Lin}=\left|\frac{T_3}{6}\phi_0^2-\frac{i\omega\tilde{\tau}}{\chi\phi_0}\right|
=\sqrt{\left(\frac{T_3}{6}\phi_0^2\right)^2+\left(\frac{\omega|\tilde{\tau}|}{\chi\phi_0}\right)^2}.
\end{equation}

Fixing the frequency to $\omega_0$ allows the optimal initial angle for a particular frequency to be found via standard calculus optimisation,

\begin{equation}
\phi_0=\left(\frac{3\sqrt{2}\omega_0|\tilde{\tau}|}{|T_3|\chi.}\right)^{\frac{1}{3}}
\end{equation}

This particular value of $\phi_0$ gives a total relative error of

\begin{equation}
\delta G^*_{Lin}=\sqrt{2\omega^2+\omega_0^2}\left(\frac{|T_3||\tilde{\tau}|^2}{12\chi^2\omega_0}\right)^{\frac{1}{3}}.
\end{equation}

Notice that this error is proportional to $|\tilde{\tau}|^{2/3}$ meaning that the error decays with the number of flips by $n^{-1/3}$. This means, at least for high frequencies, halving the relative error requires 8 times the number of flips!

%
%

\subsection{Non-linear Case}
\subsubsection{Error in Complex Shear Modulus}
Next we consider the relative error of $G^*(\omega)$ when accounting for a non-linear driving torque, as given by the analysis outlined in section \ref{sec_UnnormalisedAnalysis}. The error contribution from Brownian motion can be established by including the thermal noise term in equation \ref{eq_lagevin_nonlinear}. This yields an expression for $G^*(\omega)$,
\begin{equation}
G^*(\omega)=\frac{\chi}{\alpha}\frac{i\omega}{\phi_0-i\omega\tilde{\phi}}\left(\tilde{T}-\frac{\tilde{\tau}}{\chi}\right),
\end{equation}

\noindent with an absolute relative error of

\begin{equation}
\delta G^*_{NLin}=\left|\frac{\tilde{\tau}}{\chi\tilde{T}}\right|.
\end{equation}

Unlike the linear case, Brownian motion is the primary source of error so here the relative error is proportional to $\left|\tilde{\tau}\right|$. Hence, the relative error reduces with the number of flips at a faster rate of $n^{-1/2}$.

\subsubsection{High Frequency Error}
Applying the initial value theorem shows that at high frequencies the minimum error is obtained by maximising the initial driving torque,
\begin{equation}
\delta G^*_{NLin}=\frac{\omega|\tilde{\tau}|}{\chi T(\phi_0)}.
\end{equation}

\subsubsection{Low Frequency Error in a Viscous Fluid}

From equation \ref{eq_viscous_nonlinear}, in a viscous fluid $\tilde{T}=\frac{\phi_0}{k}$ for $\omega=0$. Therefore, the error is given by

\begin{equation}
\delta G^*_{NLin}=\frac{k|\tilde{\tau}|}{\chi \phi_0},
\end{equation}

\noindent which is minimised by maximising the initial position. These results suggest that an initial position larger than the position which maximises the driving torque should be chosen to reduce error.

%
%
%

\section{Experimental Results}

\begin{figure*}[t]
\centering
\includegraphics*[width=\textwidth]{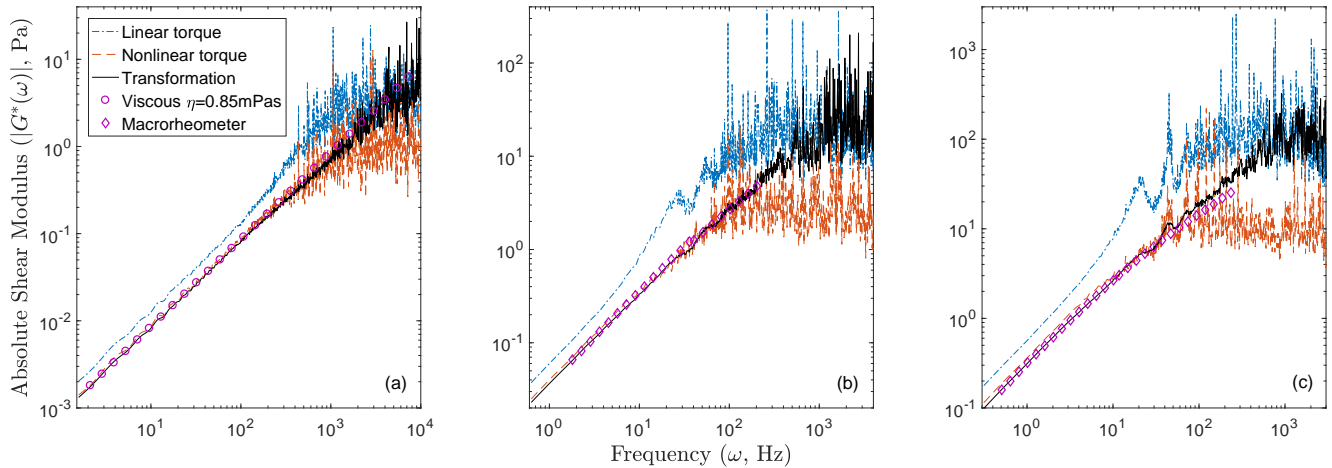}
\caption{A comparison between analysis methods in both viscous water and viscoelastic dilutions of Celluvisc eye drops. (a) depicts results of averaging 222 2s flips in water, (b) 185 5s flips in 50\% Celluvisc, and (c) 90 10s flips in 100\% Celluvisc. In each graph the blue dashed line is the shear modulus calculated using the old theory which assumes a linear restoring torque. The orange dashed lines are evaluated using the new theory accounting for the non-linear restoring torque outlined in section \ref{sec_UnnormalisedAnalysis}. The solid black line represents values obtained via the variable transformation analysis described in section \ref{sec_NormalisedAnalysis}, which mitigates error introduced by variation in initial position. All these analysis techniques are compared to either theoretical values (circles) or macrorheological measurements\cite{Bennett} (diamonds).}\label{fig_G}
\end{figure*}

Measurements of $G^*(\omega)$ were conducted in both viscous and viscoelastic fluids to compare the accuracy and precision of the new analysis methods. Applying the same methodology outlined by Zhang \textit{et al}.\cite{Shu} optical tweezers were employed to rotationally trap a spherical vaterite probe particle. The particle rotates between two stable equilibrium angles by alternating between two angularly offset linearly polarised beams.

In this case, the restoring torque function is sinusoidal $T(\phi)=1/2 \sin(2\phi)$ because of the waveplate nature of the vaterite probe particles.\cite{Nieminen2001} Therefore, the variable transformation is $\Psi=\tan\phi$ and the optimal initial angle should be within $\pi/4\leq\phi_0<\pi/2$. For measurements presented here $\phi_0\approx 70^\circ$, well beyond the linear regime.

Measurements were conducted in water, a viscous fluid as well as dilutions (50\% and 100\% by weight) of Celluvisc (Allergan) eyedrops, a strongly viscoelastic fluid. $G^*(\omega)$ of these Celluvisc dilutions has been previously measured using a macrorheometer and time-temperature superposition by Bennett \textit{et al}.\cite{Bennett}. These values, together with theoretical values of $G^*(\omega)=i\eta\omega$ in a viscous fluid can help establish the accuracy of the three different analysis methods presented in the theory section: analysis assuming a linear torque (section \ref{sec_theory_linear_case}); analysis that accounts for a non-linear torque but at the expense of normalisation (section \ref{sec_UnnormalisedAnalysis}) and finally analysis that uses a variable transformation to account for the non-linear torque and also allows normalisation (section \ref{sec_NormalisedAnalysis}).

The results, illustrated in figure \ref{fig_G}, quite clearly demonstrate the differences in accuracy and precision of the three different analysis methods in all three fluids. The method that assumed a linear torque increased the apparent shear modulus by almost a factor of 2. This is likely because the actual torque at larger angles is much less than supposed when assuming a linear torque function. Hence, the apparent viscoelasticity is larger to compensate.

Both of the other two analysis methods which account for the non-linear torque function produce values of $|G^*(\omega)|$ that have very good agreement with each other and the previous macrorheological measurements. However, the transformation method is more precise and resolves an additional decade before high frequency noise dominates the signal. Interestingly, this good agreement suggests that the flips did decay slowly relative to the fluid memory function validating the the approximation in equation \ref{eq_slow_approximation}.

\begin{figure*}[t]
\centering
\includegraphics[width=\textwidth]{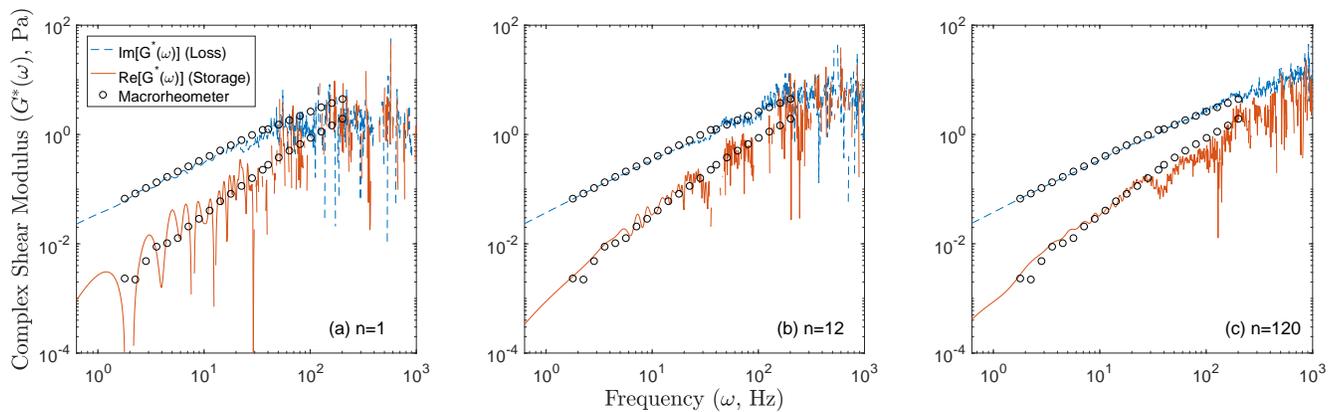}
\caption{The relationship between precision of $G^*(\omega)$ and the number of averaged flips in 50\% Celluvisc. (a) shows the viscoelasticity obtained by analysing a single 5s flip with the new method. (b) depicts results from 12 flips during a 1 minute measurement and (c) 120 flips during a 10 minute measurement. All three graphs show good agreement between the microrheological results (lines) and macrorheological data \cite{Bennett} (circles). Evidently, the precision increases with the number of averaged flips, however, because of the large amplitude of each flip, precise results can be obtained within 1 minute.}\label{fig_Num_Flips}
\end{figure*}

There are concerns about the applicability of particle tracking microrheometers inside slowly changing systems because of the long times required to obtain statistically significant averages \cite{Tassieri2015}. As depicted in figure \ref{fig_Num_Flips}, our results demonstrate that this new theory improves the signal of each flip enough to enable precise measurements of $G^*(\omega)$ in sub-minute time scales.

The signal to noise ratio of only a single 5s flip is sufficient to characterise the viscoelasticity at lower frequencies. The presence of absolute random error does, however, affect the elastic measurements more greatly because of its larger relative size. 12 flips greatly reduces random noise allowing precise measurements of both viscosity and elasticity within 1 minute. Spending 10 minutes to average 120 flips does further improve the precision with diminished returns. Therefore, this new theory endows active particle tracking microrheometers with the speed necessary to explore slowly changing biological systems that were previously inaccessible.

\section{Conclusion}

Active microrheology where a probe is impulsively driven switching between two states (two positions for translational microrheology and two orientations for rotational microrheology) can be performed with greatly improved signal to noise ratios by having larger distances or angles between the two positions or orientations. In many cases, such as where optical forces or torques are used to drive the particle, this will be outside the regime where the force or torque can be accurately approximated as a linear spring. This necessitated the development of a more general theory, not assuming linear forces.

We have presented this theory here, and shown the improvements in signal to noise that can be achieved. In addition, for some classes of problems, it is possible to further reduce error by applying a variable transformation which linearises the equation of motion. This allows normalisation that eliminates error introduced by low frequency drift in the particle's equilibrium position. Our measurements suggest that eliminating error can resolve viscoelasticity at an additional decade for higher frequencies. These improvements in the signal to noise ratio gives a significant reduction in the measurement time for a given error. Thus the method is more conducive to measuring viscoelasticity in slowly changing microscopic systems, such as a living cell.

\begin{acknowledgments}
This research was supported under Australian Research Council's
Discovery Projects funding scheme (project number DP140100753) as well as an Australian Government Research Training Program Scholarship.
\end{acknowledgments}



\begin{table*}[t]
\caption{Appendix: List of Variable Transformations}
\begin{tabular}{|c|c|c|}
\hline
$T(\phi)$, where $\beta>0$                   & $\Psi(\phi)$                                                                                                                    & Optimal $\phi_0$                                            \\ \hline
\rowcolor[HTML]{EFEFEF} 
$\frac{1}{\beta}\sin\beta\phi$               & $\frac{2}{\beta}\tan\left(\frac{\beta}{2}\phi\right)$                                                                           & $\frac{\pi}{2\beta}\leq\phi<\frac{\pi}{\beta}$           \\
$\frac{1}{\beta}\tan\beta\phi$               & $\frac{1}{\beta}\sin\beta\phi$                                                                                                  & $\phi_0=\frac{\pi}{2\beta}$                                 \\
\rowcolor[HTML]{EFEFEF} 
$\frac{1}{\beta}\sinh \beta\phi $            & $\frac{2}{\beta}\tanh \left(\frac{\beta}{2}\phi\right)$                                                                         & $\phi_0\gg 0$                                               \\
$\frac{1}{\beta}\tanh\beta\phi$              & $\frac{1}{\beta}\sinh(\beta\phi)$                                                                                               & $\phi_0\gg 0$                                               \\
\rowcolor[HTML]{EFEFEF} 
$\phi+\beta\phi^3$                           & $\frac{\phi}{\sqrt{1+\beta \phi^2}}$                                                                                            & $\phi_0\gg 0$                                               \\
$\phi-\beta\phi^3$                           & $\frac{\phi}{\sqrt{1-\beta \phi^2}}$                                                                                            & $\frac{1}{\sqrt{3\beta}}\leq\phi<\frac{1}{\sqrt{\beta}}$ \\
\rowcolor[HTML]{EFEFEF} 
$\frac{\phi+\beta\phi ^3}{1+3 \beta\phi ^2}$ & $\phi+\beta\phi^3$                                                                                                              & $\phi_0\gg 0$                                               \\
$\frac{\phi-\beta\phi ^3}{1-3 \beta\phi ^2}$ & $\phi-\beta\phi^3$                                                                                                              & $\phi_0=\frac{1}{\sqrt{3\beta}}$                            \\
\rowcolor[HTML]{EFEFEF} 
$\phi e^{-\beta\phi^2}$                      & $\frac{\text{sign}(\phi)}{\sqrt{\beta}}\text{exp}\left[\frac{1}{2}\left(\text{Ei}\left(\beta\phi^2\right)-\gamma\right)\right]$ & $\phi\geq\frac{1}{\sqrt{2\beta}}$                           \\
$\frac{\phi }{1-2 \beta\phi^2}$              & $\phi e^{-\beta\phi^2}$                                                                                                         & $\phi=\frac{1}{\sqrt{2\beta}}$                           \\ \hline
\multicolumn{3}{|c|}{$\text{Ei}(z)$ is the exponential integral function and $\gamma\approx 0.5772$ is Euler's constant.}                                                                                                                    \\ \hline
\end{tabular}
\end{table*}


\providecommand{\noopsort}[1]{}\providecommand{\singleletter}[1]{#1}%
%

\end{document}